# Graphane as polyhydride of graphene.
# Computational synthesis applied to two-side membrane

**Elena F Sheka and Nadezhda A Popova**
Peoples' Friendship University of Russia
Miklukho-Maklaya 6, Moscow 117198, Russia

E-mail: sheka@icp.ac.ru

**Abstract.** A great efficacy of molecular quantum chemistry applied to basic graphene problems has been recently demonstrated by the authors when studying the formation of peculiar composites between carbon nanotubes and graphene as well as considering tensile deformation and fracture of a graphene sheet in due course of a mechanochemical reaction. The optimistic results obtained in the studies make it possible to shift attention from the solid state problems and consider the graphane formation as multistep hydrogenation of the pristine molecule. To proceed we have to answer the following questions: 1) which kind of the hydrogen adsorption, namely, molecular or atomic, is the most probable; 2) what is a characteristic image of the hydrogen atom attachment to the substrate; 3) which carbon atom (or atoms) is the first target subjected to the hydrogen attachment and how carbon atoms are selected for the next steps of the adsorption; 4) is there any connection between the sequential adsorption pattern and cyclohexane conformers. First results obtained on this way are presented in the current paper. The calculations were performed within the framework of unrestricted broken symmetry Hartree-Fock approach by using semiempirical AM1 technique.

## 1. Introduction
On the background of ~2000 publications devoted to graphene in 2010 [1, 2], a few tens of 2010' papers [3-27], related to graphane either directly or just mentioning the subject, take their well designed niche. [1] The publication list occurs quite peculiar since only one of the papers concerns experimental study [3] while the other are related to different theoretical and/or computational aspects of the graphane science. This finding clearly highlights that in spite of the fact that the discussion of hydrogen-modified single graphite layer has a long history (see [28] and references therein) and that since 2009 the discussions have been transferred from theory and computations to the real subject [29] which caused a drastic growth of interest of researches in multiple fields, there have still been a large number of unresolved and tiny problems that need highlighting and solution. On the other hand, the list is quite representative to elucidate the state of art of the current graphane science. That is why we shall limit ourselves by the analysis of the 2010' publication first when discussing the risen problems.

---

[1] Obviously, the authors might be unable to get a complete list of graphane publications hidden under an avalanche-like stream of the issues and apply for excuse and understanding.

To facilitate the consideration of the problems as well as suggested approaches to their consideration, let us conditionally cluster the studies into three groups, namely: theoretical, computational, and molecular.

The theoretical studies concern mainly solid state aspects of graphene (see exhausted review [30] on a theoretical perspective of graphene) related to the influence of the chemical modification of pristine graphene on peculiar properties of the latter as two-dimensional crystal. Applying to hydrogenation, this concerns the effect on bandgap opening induced by patterned hydrogen adsorption [4], magnetic properties in graphene-graphane superlattices [5, 6], high-temperature electron-phonon superconductivity [7, 8], Bose-Einstein condensation of excitons [9], giant Faraday rotation [10], and a possible creation of quantum dots as vacancy clusters in the graphane body [11]. Mechanical properties of the new solid should be mentioned as well [12-15].

The computational approach to the problems in the dominant majority of papers [3-27] is based on DFT computational scheme while a few studies were performed by using molecular dynamics [15-17]. The DFT schemes applied were of different configurations but of a common characteristic: the calculations, by the only exclusion [21], have been performed using closed-shell approach (a restricted computational scheme by other words) without taking into account effectively unpaired electrons of pristine graphene at either non-completed graphene hydrogenation or at the consideration of cluster graphene vacancies inside graphane. At the same time the effective unpairing of odd electrons in graphene play a governing role in the behavior of its electronic system [31-34]. The molecular studies concern the way of the simulation of a studied object by a particular structural model. Since the above mentioned DFT studies are aimed mainly at the solid state aspects of graphane crystal, they are based on supercell presentation followed by periodic boundary conditions (PBC). In the majority of cases, the supercell is just unit cell of the chairlike-conformed graphane crystal taken for granted. The unit cells of more complicated structure, predetermined beforehand, are exploited as well in the studies devoted to the elucidation of peculiarities of the electronic system of graphane itself and of its physico-chemical properties [15, 18- 23]. However, if one concentrates on the latter, between which the chemical reaction of graphene hydrogenation is the main topic, a presentation of graphane as a "giant molecule" firstly proposed in the fundamental paper [29] and then supported by chemists (see [3] and a comprehensive review [25]) occurs very perspective since it makes possible to support on large facilities of quantum chemistry in studying molecules. Among the discussed 2010' papers the approach was used, at least partially, in [23] and [24].

A great efficacy of molecular quantum chemistry applied to basic graphene problems has been recently demonstrated by the authors when studying the formation of peculiar composites between carbon nanotubes and graphene [35] (E.Sh.) as well as considering tensile deformation and fracture of a graphene sheet in due course of a mechanic-chemical reaction [36-38] (E.Sh. and N.P.). The optimistic results obtained in the studies make it possible to shift attention from the solid state problems and to get answers to the following questions: 1) what is the reaction of the graphene hydrogenation, 2) which kind of final products and 3) at which conditions the formation of the graphene polyhydrides can be expected. First results obtained on this way are presented in the current paper.

## 2. Algorithmic computational synthesis of graphene polyhydride $(CH)_n$

From the chemical viewpoint the graphene hydrogenation results from the hydrogen adsorption on the basal plane of either graphite or graphene. Both structural configurations are practically equivalent from the computational viewpoint due to weak interaction between neighboring graphite layers.

According to this, computational studying of the hydrogen adsorption on graphite was restricted to the consideration of either one or two layers of substrates [28, 39-41]. That is why the graphene hydrogenation had been actually studied long before the material was obtained. During these studies has become clear that the following items should be elucidated when designing the computational problem:

1. Which kind of the hydrogen adsorption, namely, molecular or atomic, is the most probable?
2. What is a characteristic image of the hydrogen atom attachment to the substrate?
3. Which carbon atom (or atoms) is the first target subjected to the hydrogen attachment?
4. How carbon atoms are selected for the next steps of the adsorption?
5. Is there any connection between the sequential adsorption pattern and cyclohexane conformers?

Experimentally is known that molecular adsorption of hydrogen on graphite is extremely weak so that its hydrogenation occurs only in plasma containing atomic hydrogen. No computational consideration of the molecular adsorption has been so far known as well.

In the first paper devoted to DFT study of a single hydrogen atom adsorption on the graphite substrate [39] was shown that the hydrogen was accommodated on-top carbon atom forming a characteristic C-H bond of 1.11A in length, and caused a significant deformation of the substrate due to the transformation of $sp^2$ configuration of the pristine carbon atom into $sp^3$ one after adsorption. Each of these issues is highly characteristic for individual adsorption event and significantly influences the formation of final product.

The site dependence of adsorption events on graphite substrate was studied in the framework of classical supercell-PBC DFT scheme by using restricted (RDFT) [28] and unrestricted (UDFT) [40] computational scheme. Positions for the first adsorption act were taken arbitrarily while sites of the subsequent hydrogen deposition were chosen following the lowest-energy criterion when going from one atom of the supercell to the other. If only two hydrogen atoms were positioned in [40], eight atoms were accommodated within 3x3x1 supercell in [28] for which 46 individual positions were considered. It is worthwhile to note that study [28] should be related to the first one concerning graphene since the substrate was presented by a single graphite layer. At the same time, attributing their study to graphite, the authors restricted themselves by one-side adsorption due to inaccessibility of the other side of the layer to hydrogen gas in the graphite bulk. Relating their study to graphene, authors of Ref. 42 performed classical molecular dynamics calculations studying the accommodation of an individual hydrogen atom and their pair on clusters consisting of 50 and 160 carbon atoms, respectively. Their results were quite similar to those obtained for graphite [39]. The choice of the best deposition site was subordinated to the lowest-energy criterion as well. These studies showed that the hydrogen adsorption on either graphite or graphene basal plane is possible but the final product does not present any new material and is just a chemically modified graphite and/or graphene patterned by hydrogen adsorbate.

For the first time the topic concerning the formation of new material in due course of chemical modification of graphite was concerned with the formation of graphite polyfluoride $(CF)_n$[2] (see a comprehensive review in [43]) for which Charlier et all suggested well defined regular structure supported computationally [44]. The authors followed the Rüdorff suggestion presenting the fluoride structure as an infinite array of *trans*-linked perfluorocyclohexane chairs [43]. The Rüdorff structure was based on the existence of two stable conformers of cyclohexane known as chairlike and boatlike configurations (see Wikipedia). The former conformer is more energetically favorable but commercial product always contains a mixture of the two. As for $(CF)_n$, there was an evidence of the structure as an infinite array of *cis-trans*-linked cyclohexane boats [45] so that Charlier et all included in their study both conformations and showed that the preference should be given to the chairlike one. In view of intensified study of the graphite hydrogenation and of a large resonance, which achieved its maximum by 2004, caused by a massive examination of both fluorinated and hydrogenated polyderivatives of fullerene $C_{60}$ that exhibited a deep parallelism between these two $sp^2$-carboneous families [46, 47], appearing the paper of Sofo et all. [48] was absolutely motivated. The authors led the parallelism of fluorinated and hydrogenated $sp^2$ fullerenes whilst not referring to these fundamental studies, in the foundation of their approach and followed the same scheme of the construction and computational treatment of graphene polyhydride $(CH)_n$ that were discussed for polyfluorides $(CF)_n$ by Charlier et all. [44]. Thus introduced new substance is known now as graphane. Similarly to $(CF)_n$, the chairlike conformer $(CH)_n$ occurred more energetically stable which was later confirmed by studies of another conformers additionally to the basic chairlike and boatlike, namely, tablelike [18], 'zigzag' and 'arm chair' configurations [19], and 'stirrup' one [20]. Among these structural modes, the chairlike one has obtained the largest acceptance when discussing the influence of the hydrogenation of graphene on its unique properties as 2D-crystal that was discussed in Introduction.

---

[2] $(CH)_n$ and $(CF)_n$ species are referred to in the literature as monohydride and monofluoride graphene, respectively. The notation was attributed to crystals with unit cells containing either CH or CF units. Addressing molecular approach, it would be better to consider the configuration $(CH)_n$ and $(CF)_n$ as polyderivatives of graphene, which makes it possible to consider their formation in due course of a stepwise hydrogenation and/or fluorination.

In spite of seemingly exhaustive exploration of the hydrogenated graphene structure, there have been so far problems that were not elucidated during these studies. The first concerns the confirmation of the fluoro-hydrogen parallelism in regards to polyfluorides and polyhydrides of graphene. The second is related to the exhibition of the hydrogenation process that has evidently to proceed via the formation of cyclohexane conformers of graphane. The third raises the question at which conditions the chairlike conformer can dominate. To answer the questions, we have to get at hands a definite way of a controlled exhibition of the subsequent hydrogenation of the pristine graphene. To achieve the aim we suggest to present a piece of graphene as a 'giant' molecule and to apply the approach of a computational synthesis of molecular polyderivatives that was used in the case of fluorinated [49, 50] and hydrogenated [51, 52] fullerene $C_{60}$ (see a comprehensive review of the approach in the coming monograph [53]).

The computational synthesis of polyderivatives of $sp^2$ nanocarbons is based on the peculiarities of odd-electron structure of fullerenes, nanotubes, and graphene [32, 53, 54]. Due to exceeding C-C bond length the critical value of 1.395Å, under which a complete covalent bonding between odd electrons thus transferring them into classical π electrons occurs only, the odd electrons become partially unpaired while the species obtain a partially radical character. The unrestricted broken symmetry approach makes it possible to evaluate the total number of the unpaired electrons $N_D$ as well as their partial distribution over atoms $N_{DA}$ that quantify the molecular chemical susceptibility (MCS) and atomic chemical susceptibility (ACS), respectively [53]. Distributed over molecule atoms, ACS forms a chemical activity map of the molecule exhibiting sites from the highest chemical activity (large ACS values) to the lowest ones (small ACS values) thus designing 'a chemical portrait' of the molecule [54]. The highest ACS value (high-rank ACS) points to target atom that first enters the chemical reaction. After the first chemical attack is completed and the first derivative is constructed, the ACS map of the latter highlights the target atom(s) for the next chemical addition so that the second derivative is constructed. The ACS map of this derivative opens the site for the next attack, and so forth, so that the algorithmic stepwise manner of a consequent construction forms the grounds of a controlled computational synthesis of the polyderivatives guided by the highest ACS quantities. The approach has exhibited its efficacy applying to the mentioned polyfluorides and polyhydrides families of $C_{60}$ [49-52], as well as to the fullerene polycyanides and polyaziridines [55], polyamines of complicated structure [53], and graphene-carbon nanotubes composites [35].

Following this algorithm, we have performed study of the stepwise computational synthesis of polyhydrides $(CH)_n$ by addition of either hydrogen molecule or atomic hydrogen to the graphene molecule. A similar study concerning polyfluorides $(CF)_n$ is in progress. The molecule is considered a membrane, whose atoms were accessible to adsorbate either from both sides or from only one. The membrane, in its turn, is either strictly fixed over its perimeter (fixed membrane) or its edge atoms are allowed to move freely under optimization procedure (free standing membrane). The main results, detailed description of which will be published elsewhere, are the following.
1. Molecular adsorption of hydrogen on the graphene is weak and preferable to the one-side free standing membrane. It is accompanied with small coupling energy and low saturation covering.
2. Atomic adsorption is strong, the saturation covering approaches and/or achieves 100%, the adsorption pattern is of complicated mixture of different cyclohexane conformational motives, the prevalence of which depends on the membrane state and on the accessibility of the membrane sides.
    a. The 100% hydrogen covering of the fixed membrane with the both sides accessibility forms a regular structure consisting of trans-coupled chairlike cyclohexane conformers thus revealing chairlike graphane.
    b. The hydrogen covering of the fixed membrane with one-side accessibility for hydrogen atoms forms a continuous accommodation of tablelike cyclohexane conformers over convex surface similar to the exterior image of a delta plane canopy.
    c. The hydrogen covering of the free standing membrane with the both sides accessibility exhibits an initiated rolling of the pristine graphene plane, caused by the admixture of the boatlike cyclohexane conformers to the regular motive of chairlike structures.
    d. The hydrogen covering of the free standing membrane with one-side accessibility for hydrogen atoms causes so significant rolling of the pristine graphene plane, that the conformer analysis of the final configuration is difficult.

The current paper presents in details the results related to the formation of the regular graphane structure under hydrogen atom two-side adsorption onto the fixed membrane. This very case can be attributed to the real experimental conditions under which the appearance of the graphane as a new 2D-crystal has been heralded [29]. The amorphous graphane disclosed experimentally was attributed by authors of ref. 29 to the one-side adsorption on the ripples formed by graphene deposited on the $SiO_2$ substrate. This case is well suited to case *b* of our study.

Calculations were performed in the framework of the unrestricted broken symmetry Hartree-Fock (UBS HF) scheme implemented in semiempirical AM1 version of the CLUSTER-Z1 codes [56].

## 3. The first stage of the graphene hydrogenation

The graphene molecule selected for the study presents a (5,5) nanographene in terms of ($n_a$, $n_z$) presentation suggested in [57]. Numbers $n_a$ and $n_z$ count benzenoid units along armchair (*ach*) and zigzag (*zg*) edges of graphene rectangle. Equilibrium structure of the molecule alongside with its ACS map, which presents the distribution of atomically-matched effectively unpaired electrons $N_{DA}$ over graphene atoms, is shown in figure 1. Panel *b* exhibits the ACS distribution attributed to the atoms positions thus presenting the 'chemical portrait' of the graphene sheet. Different ACS values are plotted in different colors according to color scale. The absolute ACS values are shown in panel *c* according the atom numbering in the output file. This numbering will be kept through over results presented in the current paper. As seen in the figure, 22 edge atoms involving 2x5 *zg* and 2x6 *ach* ones have the highest ACS thus marking the perimeter as the most active chemical space of the molecule. The maximum absolute ACS values of 1.3*e* and 1.1*e* on *zg* and *ach* atoms, respectively, demonstrate a high radicalization of these atoms and obviously select these atoms out of the other ones as first targets for hydrogenation. It should be noted that the discussed ACS values are directly connected with spin density on edge atoms (see a detail discussion of the topic in [31-34]), still actively discussed for about twenty years [17] since the first publication [58] in 1996. The appearance of the density itself is just a witness of the spin contamination of the solution obtained for singlet-ground-state graphene (sheets, ribbons, molecules, quantum dots, etc.) by using single-determinant quantum schemes of either UBS DFT or UBS HF approaches [59, 31-34]. In the frameworks of these computational schemes, spin density, which is extra for singlet ground state, evidences directly partial or complete unpairing of weakly interacting odd electrons but had nothing with magnetic susceptibility of the object. The edge atoms are highlighted by the fact that each of them posses two odd electrons, the interaction between which is obviously weaker than that for the basal atoms due to which the extent of the electron unpairing is the highest for these atoms, therewith bigger for *zg* edges in contrast to *ach* ones.

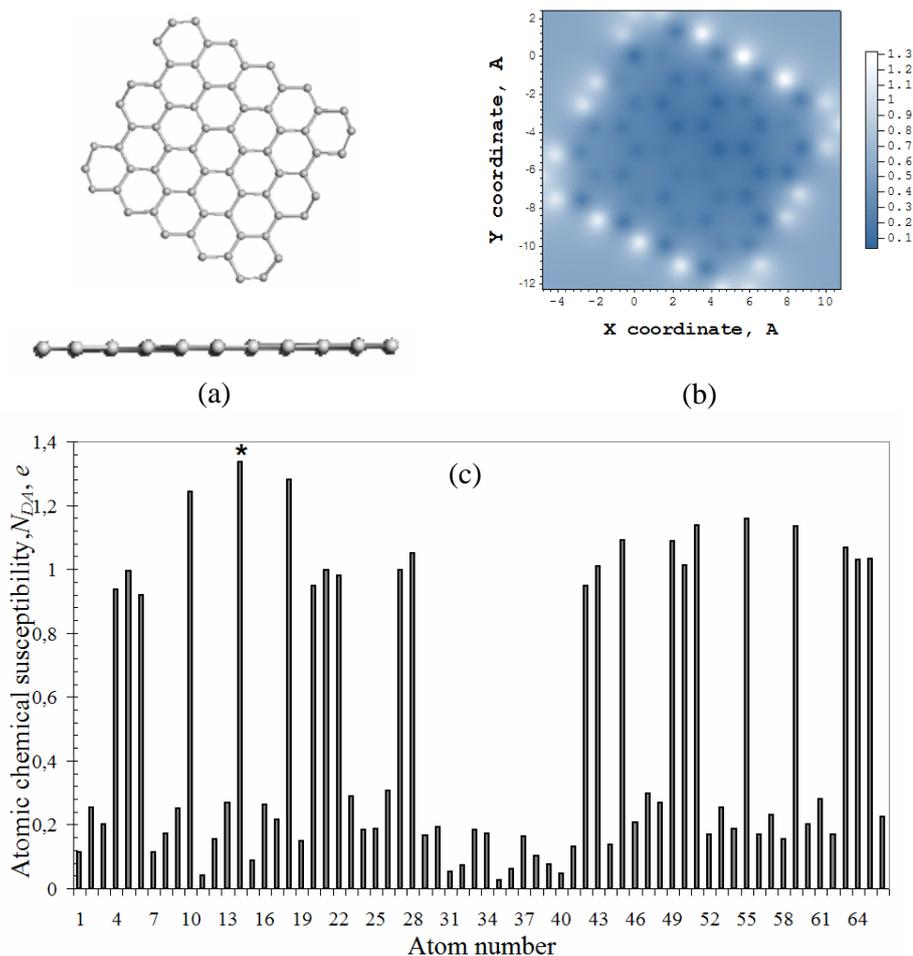

**Figure 1**. Top and side views of the equilibrium structure of (5,5) nanographene (a) and ACS distribution over atoms in real space (b) and according to atom number in the output file.

The hydrogenation of the (5, 5) molecule will start on atom 14 (star marked in fig.1c) with the highest rank ACS in the output file. As turned out, the next step of the reaction involves the atom from the edge set as well and this continues until all edge atoms are saturated by a pair of hydrogen atoms each since all 44 steps are accompanied by the high rank ACS list where edge atoms take the first place. Thus obtained the hydrogen-framed graphene molecule is shown in figure 2 alongside with the corresponding ACS map. Two equilibrium structures are presented. The structure in panel *a* corresponds to a completed optimization of the molecule structure without any restriction. Afterwards positions of edge carbon atoms and framing hydrogen atoms have been fixed and the optimization procedure was repeated leading to the structure shown in panel *c*. In what follows, we shall refer to the two structures as free standing and fixed membranes. Blue atoms in fig. 2d alongside with framing hydrogens are excluded from the forthcoming optimization under all steps of the further hydrogenation.

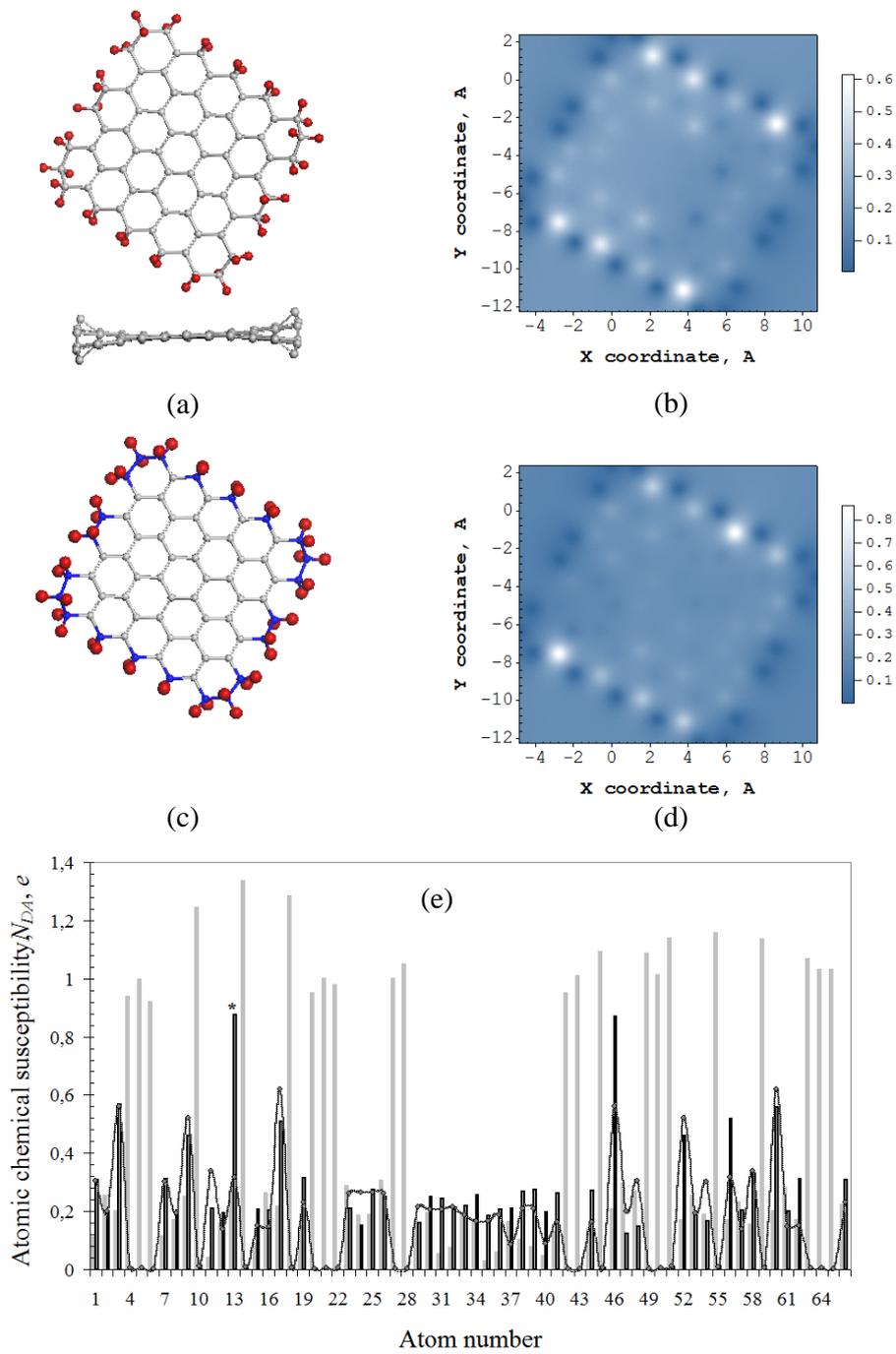

**Figure 2**. Equilibrium structures of free standing (top and side views) (a) and fixed (b) (5,5) graphene membrane and ACS distribution over atoms in real space (b, d) and according to the atom number in output file (e). Light gray histogram plots ACS data for the pristine graphene. Curve with dots and black histogram are related to membranes in panels *a* and *c*, respectively.

Chemical portraits of the structures shown in figure 2b and figure 2d are quite similar and reveal the transformation of brightly shining edge atoms in figure 1b into dark spots. Addition of two hydrogen atoms to each of the edge ones saturates the valence of the latter completely, which results in zero ACS values, as is clearly seen in figure 2e. The chemical activity is shifted to the neighboring inner atoms and retains much more intense in the vicinity of *zg* edge atoms. Since ACS distribution synchronically reflects that one for the excess of C-C bond length in $sp^2$ structures over $R_{crit}=1.395$Å [32, 33], the difference in the details of pictures in figure 2b and figure 2d highlights the redistribution of C-C bond length of free standing membrane when it is fixed on periphery.

## 4. Hydrogenation on the basal plane

The next stage of hydrogenation concerns the basal plane of the fixed membrane shown in figure 2c. To facilitate presentation of the subsequent results, the framing hydrogen atoms will not be shown in what follows. As follows from figure 2e, the first hydrogenation step should concern basal atom 13 marked by star. Since according to our conditions, the membrane is accessible to hydrogen from both sides, we have to check which deposition of the hydrogen atom, namely, above the carbon plane (up) or below it (down) is more energetically favorable. As seen from chart 1, the up position is evidently preferential and the obtained equilibrium structure H1 is shown in figure 3. The atomic structure is accompanied with ACS map that makes it possible to trace the transformation of the chemical activity of the graphene molecule during hydrogenation. Rows HN (M) in chart 1 display intermediate graphene hydrides involving N atoms of adsorbed hydrogen where H0 is related to the pristine (5, 5) nanographene with 44 framing hydrogen atoms. M points carbon atom to which the $N^{th}$ hydrogen atom is attached. Columns $N_{DA}$ and $N_{at}$ present the high rank ACS values and number of atoms to which they belong. The calculated heat of formation ΔH serves as an energy criterion for selecting the best isomorphs that are selected out in the chart by light blue coloring.

After deposition of hydrogen atom 1 on basal atom 13, the ACS map has revealed carbon atom 46 for the next deposition (see H1 ACS map in figure 3). The energy criterion favors the down position for the hydrogen on this atom so that we obtain structure H2 shown in figure 3. The deposition of the second atom highlights next targeting atom 3 (see ACS map of H2 hydride), the third adsorbed atom activates target atom 60, the fourth does the same for atom 17, and so forth. Checking up and down deposition, a choice of the best configuration can be performed and the corresponding equilibrium structures for a selected set of hydrides from H1 to H11 are shown in figure 3. As follows from the results obtained, the first 8 hydrogen atoms are deposited on substrate atoms characterized by the largest ACS peaks in figure 2b. After saturation of these most active sites, the hydrogen adsorption starts to fill the inner part of the basal space in a rather non-regular way therewith, which can be traced in figure 3 and figure 4. And the first hexagon unit with the cyclohexane chairlike motive is formed when the number of hydrogen adsorbates achieves 38. This finding well correlates with experimental observation of a disordered, seemingly occasionally distributed, adsorbed hydrogen atoms on the graphene membrane at similar covering [60].

This computational scheme continues until the $34^{th}$ step when the ACS list exhibits only zero values as shown in chart 2. ACS going to zero is connected with a continuous redistribution of the C-C bond length over the graphene body in due hydrogenation. Leaving the detailed discussion of the nanographene structure changing caused by the $sp^2$-$sp^3$ transformation of the electronic configuration of carbon atoms due to hydrogen adsorption to the next section, let us draw attention to a considerable shortening of C-C bond length shown in figure 5 relatively to H34 hydride. Five bonds of 1.35Å in length connect the remainder 10 untouched basal $sp^2$ carbon atoms. The C-C distance is much less then even C-C bond in benzene molecule so that odd electrons of these atoms are covalently coupled by pairs forming classical π electrons without any effectively unpaired fraction. In this case we can continue to consider the hydrogen adsorption individually for every pair just checking each atom of the pair taking into account a possibility of the up and down accommodation. Thus, as follows from chart 2, for a pair connecting atoms 19 and 16 that is highlighted by the ACS list in spite of zero quantities in the output file, the preference for the 35[th] deposition should be given to atom 16. Attaching hydrogen to this atom immediately highlights the pairing atom 19 that becomes a target for the 36[th] deposition. The consideration of the highlighted pair joining atoms 54 and 29 discloses a preference towards the up deposition on atom 29, and so forth until the last atom 44 is occupied by adsorbate. The structure obtained at the end of the 44[th] step is shown at the end of figure 4. It is perfectly regular, including framing hydrogen atoms thus presenting a computationally synthesized chairlike (5, 5) nanographane.

**Chart 1.** Explication of the subsequent steps of the hydrogenation of (5, 5) nanographene. ΔH in kcal/mol.

| H0 | | H1 (13) | | H2 (46) | | H3 (3) | | H4 (60) | | H5 (17) | | H6 (52) | | H7 (9) | |
|---|---|---|---|---|---|---|---|---|---|---|---|---|---|---|---|
| $N_{at}$ | $N_{DA}$ | $N_{at}$ | $N_{DA}$ | $N_{at}$ | $N_{DA}$ | $N_{at}$ | $N_{DA}$ | $N_{at}$ | $N_{DA}$ | $N_{at}$ | $N_{DA}$ | $N_{at}$ | $N_{DA}$ | $N_{at}$ | $N_{DA}$ |
| 13 | 0,87475 | up | | up | | up | | up | | up | | up | | up | |
| 46 | 0,87034 | 46 | 0,62295 | 3 | 0,56119 | 60 | 0,55935 | 17 | 0,54497 | 52 | 0,49866 | 56 | 0,47308 | 56 | 0,47318 |
| 3 | 0,56633 | 3 | 0,61981 | 60 | 0,55965 | 17 | 0,54302 | 56 | 0,50027 | 56 | 0,49856 | 9 | 0,47072 | 12 | 0,42593 |
| 60 | 0,55862 | 60 | 0,56055 | 17 | 0,54299 | 56 | 0,51849 | 52 | 0,49996 | 9 | 0,47193 | 12 | 0,36783 | 8 | 0,42263 |
| 56 | 0,51946 | ΔH = 126.50 | | ΔH = 101.88 | | ΔH = 96.88 | | ΔH = 99.77 | | ΔH = 95.00 | | ΔH = 96.82 | | ΔH = 97.50 | |
| 17 | 0,50841 | down | | down | | down | | down | | down | | down | | down | |
| 52 | 0,45883 | 46 | 0,62197 | 3 | 0,56174 | 60 | 0,55906 | 17 | 0,54576 | 52 | 0,49684 | 9 | 0,47392 | 56 | 0,46883 |
| 9 | 0,45843 | 3 | 0,62099 | 60 | 0,55926 | 17 | 0,54279 | 56 | 0,49878 | 56 | 0,49602 | 56 | 0,47247 | 12 | 0,41783 |
| 58 | 0,33071 | 60 | 0,55979 | 17 | 0,54297 | 56 | 0,51736 | 52 | 0,49782 | 9 | 0,47141 | 12 | 0,36857 | 8 | 0,41459 |
| 19 | 0,31539 | ΔH = 127.54 | | ΔH = 98.19 | | ΔH = 101.90 | | ΔH = 95.99 | | ΔH = 99.40 | | ΔH = 96.23 | | ΔH = 98.92 | |

| H8 (56) | | H9 (12) | | H10 (57) | | H11 (53) | | H12 (58) | | H13 (33) | | H14 (30) | | H15 (34) | |
|---|---|---|---|---|---|---|---|---|---|---|---|---|---|---|---|
| $N_{at}$ | $N_{DA}$ | $N_{at}$ | $N_{DA}$ | $N_{at}$ | $N_{DA}$ | $N_{at}$ | $N_{DA}$ | $N_{at}$ | $N_{DA}$ | $N_{at}$ | $N_{DA}$ | $N_{at}$ | $N_{DA}$ | $N_{at}$ | $N_{DA}$ |
| up | | up | | up | | up | | up | | up | | up | | up | |
| 12 | 0,42462 | 57 | 0,42824 | 53 | 0,30873 | 58 | 0,54014 | 33 | 0,33339 | 30 | 0,40225 | 34 | 0,34848 | 35 | 0,51203 |
| 8 | 0,42404 | 53 | 0,42131 | 8 | 0,30851 | 30 | 0,32126 | 8 | 0,30396 | 34 | 0,3618 | 31 | 0,33662 | 31 | 0,36904 |
| 53 | 0,41554 | 47 | 0,30877 | 2 | 0,2956 | 8 | 0,32075 | 2 | 0,29194 | 8 | 0,34811 | 8 | 0,2934 | 2 | 0,29541 |
| ΔH = 99.83 | | ΔH = 118.04 | | ΔH = 79.46 | | ΔH = 74.17 | | ΔH = 73.81 | | ΔH = 65.87 | | ΔH = 70.90 | | ΔH = 73.11 | |
| down | | down | | down | | down | | down | | down | | down | | down | |
| 12 | 0,4232 | 57 | 0,42759 | 8 | 0,30801 | 58 | 0,58671 | 33 | 0,32811 | 30 | 0,40175 | 34 | 0,34603 | 35 | 0,5002 |
| 57 | 0,42303 | 53 | 0,4216 | 2 | 0,29368 | 8 | 0,32092 | 8 | 0,30638 | 34 | 0,3625 | 31 | 0,3272 | 31 | 0,36752 |
| 8 | 0,42294 | 47 | 0,31082 | 53 | 0,29305 | 30 | 0,30986 | 2 | 0,28938 | 8 | 0,34298 | 2 | 0,29516 | 2 | 0,29455 |
| ΔH = 98.98 | | ΔH = 89.19 | | ΔH = 107.96 | | ΔH = 101.20 | | ΔH = 68.89 | | ΔH = 84.89 | | ΔH = 67.58 | | ΔH = 71.83 | |

## 5. Hydrogenation-induced (5, 5) nanographene structure transformation

Stepwise hydrogenation is followed by the gradual substitution of $sp^2$-configured carbon atoms by $sp^3$ ones. Since both valence angles between the corresponding C-C bonds and the bond lengths are noticeably different in the two cases, the structure of the pristine nanographene becomes pronouncedly distorted. The formation of chairlike hexagon pattern is another reason for a serious violation of the pristine close-to-plane structure. The necessary changing in the basal plane structure is vividly seen in figure 4.

Figure 5 demonstrates the transformation of the molecule structure in due course of hydrogenation exemplified by changes within a fixed set of C-C bonds. Light gray curve shifted up by 0.15Å at each panel presents the reference bond length distribution for the pristine nanographene framed by 44 hydrogen atoms that saturate 'dangling bonds' of all edge atoms of the graphene. As seen in the figure, the first steps of hydrogenation are followed by the elongation of C-C bonds that involve not only newly formed $sp^3$ atoms but some of $sp^2$ ones as well.

Comparing the pristine diagram with those belonging to a current hydrogenated species makes it possible to trace the molecule structure changes. As might be naturally expected, the $sp^2$-$sp^3$ transformation causes the appearance of elongated C-C bonds, the number of which increases when hydrogenation proceeds. To keep the carbon skeleton structure closed and to minimize the deformation caused by the structure distortion, some C-C bonds shorten their length as was already mentioned. However, when a complete hydrogenation is achieved a perfectly regular structure is obtained in contrast to quite non-regular structure of pristine framed nanographene.

**Chart 2**. A conclusive stage of the hydrogenation of (5, 5) nanographene

| H33 (66) | | H34 (41) | | H35 (16) | | H36 (19) | | H37 (29) | | H38 (54) | |
|---|---|---|---|---|---|---|---|---|---|---|---|
| $N_{at}$ | $N_{DA}$ | $N_{at}$ | $N_{DA}$ | $N_{at}$ | $N_{DA}$ | $N_{at}$ | $N_{DA}$ | $N_{at}$ | $N_{DA}$ | $N_{at}$ | $N_{DA}$ |
| up | | up | | up | | up | | up | | up | |
| 41 | 0,87719 | 19 | 0,00011 | 19 | 0,87268 | 54 | 0 | 54 | 0,88406 | 37 | 0 |
| 143 | 0,03036 | 16 | 0,00011 | 145 | 0,03623 | 29 | 0 | 147 | 0,02945 | 62 | 0 |
| 137 | 0,02556 | 115 | 0 | 134 | 0,02774 | 7 | 0 | 121 | 0,02916 | 120 | 0 |
| $\Delta H = 17.51$ | | $\Delta H = -20.25$ | | $\Delta H = 1.45$ | | $\Delta H = -52.$ Kcal/mo | | $\Delta H = -118.95$ | | $\Delta H = -115.63$ | |
| down | | down | | down | | down | | down | | down | |
| 41 | 0,87536 | 19 | 0 | 19 | 0,88234 | 54 | 0 | 54 | 0,8823 | 37 | 0 |
| 85 | 0,03146 | 16 | 0 | 107 | 0,0311 | 29 | 0 | 147 | 0,03469 | 62 | 0 |
| 137 | 0,03045 | 115 | 0 | 134 | 0,02987 | 7 | 0 | 121 | 0,02329 | 137 | 0 |
| $\Delta H = 6.36$ | | $\Delta H = 19.00$ | | $\Delta H = -27.22$ | | $\Delta H = -19.03$ | | $\Delta H = -105.70$ | | $\Delta H = -147.29$ | |
| H39 (37) | | H40 (62) | | H41 (32) | | H42 (7) | | H43 (36) | | H44 (11) | |
| $N_{at}$ | $N_{DA}$ | $N_{at}$ | $N_{DA}$ | $N_{at}$ | $N_{DA}$ | $N_{at}$ | $N_{DA}$ | $N_{at}$ | $N_{DA}$ | $N_{at}$ | $N_{DA}$ |
| up | | up | | up | | up | | up | | up | |
| 62 | 0,88414 | 7 | 0 | 7 | 0,88308 | 11 | 0 | 11 | 0,88222 | 53 | 0 |
| 149 | 0,02943 | 32 | 0 | 151 | 0,03496 | 36 | 0 | 153 | 0,0351 | 1 | 0 |
| 120 | 0,02886 | 11 | 0 | 135 | 0,02033 | 126 | 0 | 119 | 0,02231 | 128 | 0 |
| $\Delta H = -147.15$ | | $\Delta H = -144.19$ | | $\Delta H = -162.81$ | | $\Delta H = -201.06$ | | $\Delta H = -187.39$ | | $\Delta H = -227.61$ | |
| down | | down | | down | | down | | down | | down | |
| 62 | 0,88269 | 7 | 0 | 7 | 0,88315 | 11 | 0 | 11 | 0,88403 | 48 | 0 |
| 149 | 0,0346 | 32 | 0 | 151 | 0,03071 | 36 | 0 | 153 | 0,02955 | 20 | 0 |
| 120 | 0,02261 | 11 | 0 | 127 | 0,02866 | 136 | 0 | 119 | 0,02884 | 23 | 0 |
| $\Delta H = -134.71$ | | $\Delta H = -175.29$ | | $\Delta H = -173.52$ | | $\Delta H = -170.64$ | | $\Delta H = -200.10$ | | $\Delta H = -198.25$ | |

The hydrogen atom behavior can be characterized by the average length of C-H bonds that they form. As seen in figure 5a, 44 newly formed bonds are quite uniform with the average value of 1.125Å. The value slightly exceeds the accepted C-H length for hydrocarbons at the level of 1.10-1.11Å. This is not an evidence of the computational program inconsistency since the UBS HF calculations within AM1 version provide the C-H of 1.10Å in length for a lot of hydrocarbons including the aromatic family from benzene to pentacene as well as for H-terminated carbon nanotubes of [61]. But the same computational scheme provides longer C-H bonds of 1.127Å for fullerene hydrides [51, 52] and H-terminated graphene, both monoatomic- [31-33] and diatomic-terminated. In the latter case, the matter is about peripheral framing conserving empty basal plane. As soon as the hydrogenation starts to proceed on this place, the length of C-H bond rapidly grows up to 1.152 Å for H8 (figure 5a), 1.139 Å for H18 (figure 5b), 1.133 Å for H34 (figure 5c), but comes back to the framing average length of 1.126 Å when regular graphane structure is completed (figure 5d). A similar situation was observed in the case of fullerene $C_{60}$ hydrides, for which a non-completed

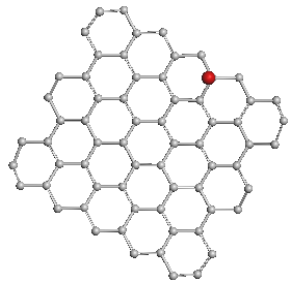 H1 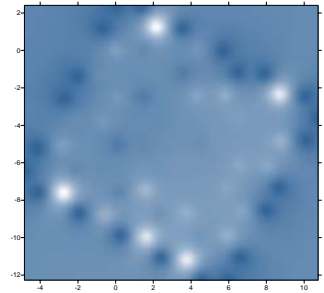

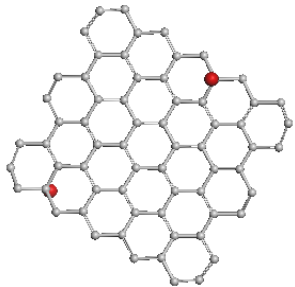 H2 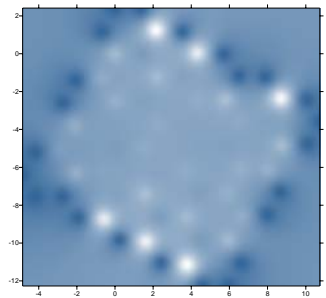

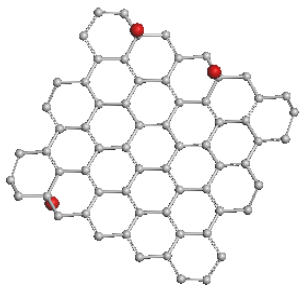 H3 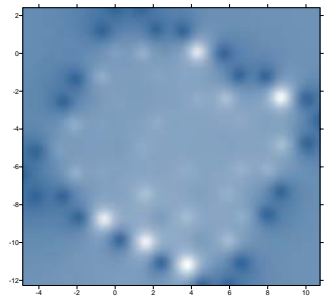

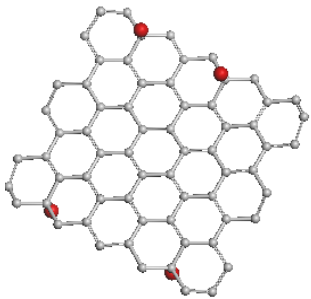 H4 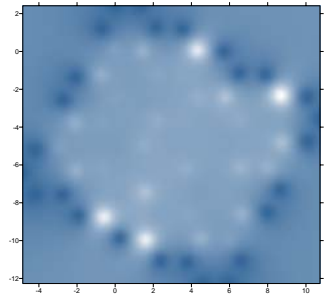

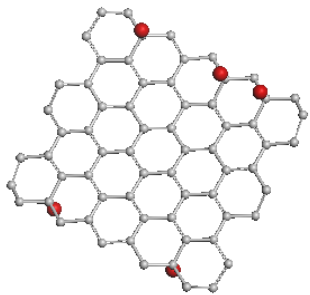 H5 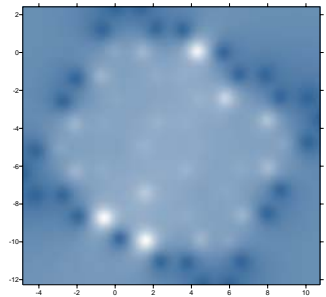

**Figure 3** contnd

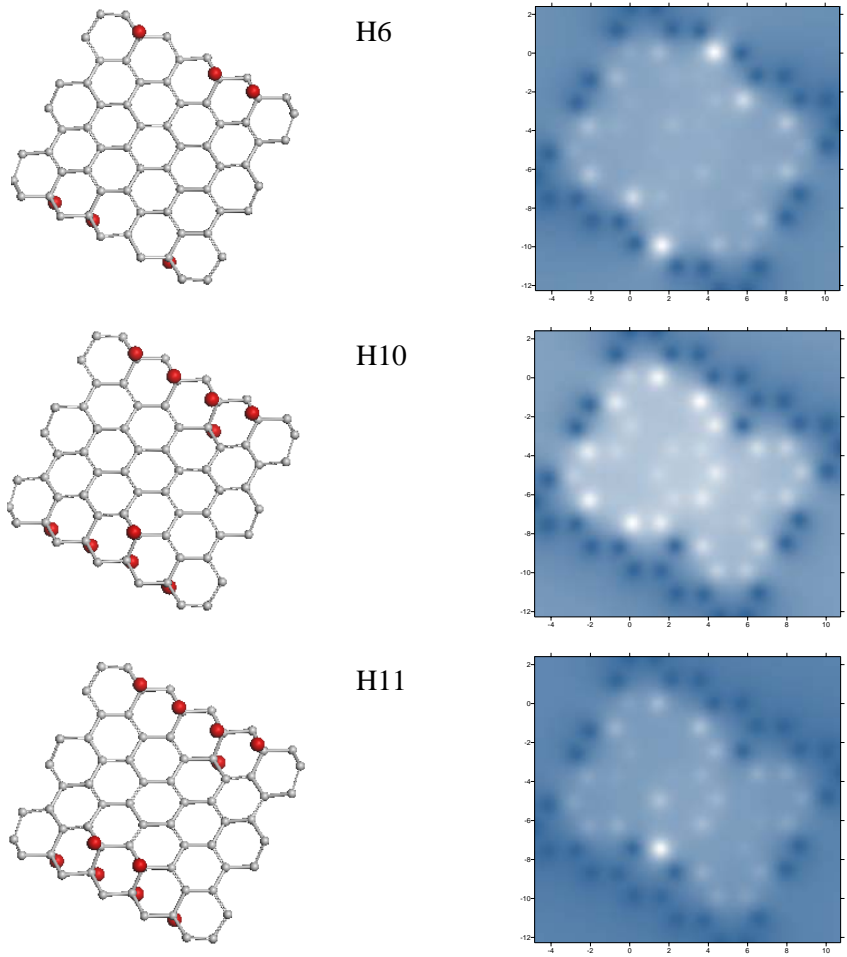

**Figure 3.** Equilibrium structure (left) and real-space ACS maps (right) of intermediate hydrides from the initial stage of the basal-plane hydrogenation.

hydrogenation is followed by elongated C-H bonds of 1.146 Å in average while the hydrogenation completion resulted in equilibration of all bonds at the level of 1.127Å in length.

## 6. Energetic characteristics accompanying the nanographene hydrogenation

A *brutto* coupling energy that may characterize the molecule hydrogenation can be presented as

$$\Delta E_{cpl}^{brutto}(n) = E_{cpl}^{tot}(n) = \Delta H_{nHgr} - \Delta H_{gr} - n\Delta H_{at}. \quad (1)$$

Here $\Delta H_{nHgr}$, $\Delta H_{gr}$, and $\Delta H_{at}$ present heats of formation of graphene hydride with *n* hydrogen atoms, a pristine nanographene, and hydrogen atom, respectively. This energy has turned out to be negative with absolute value gradually increased when the number of the attached hydrogen atoms grows. The finding makes it possible to conclude that the considered hydrogenation of (5, 5) nanographene is energetically profitable. The tempo of hydrogenation may be characterized by the coupling energy needed for the addition of each next hydrogen atom. It can be determined as

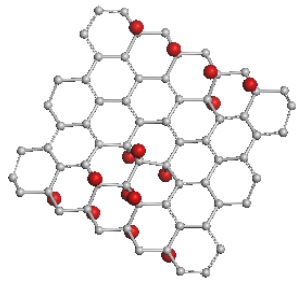 H15 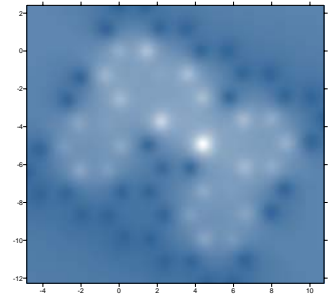

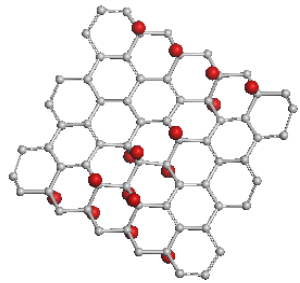 H16 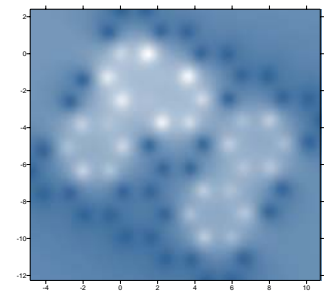

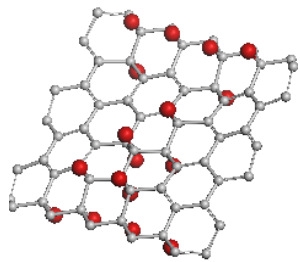 H17 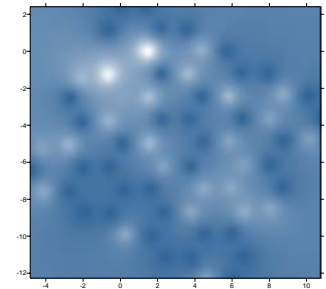

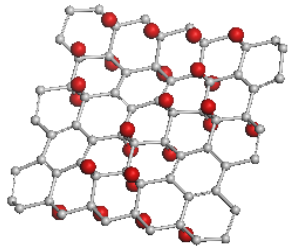 H25 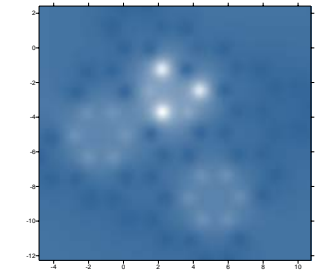

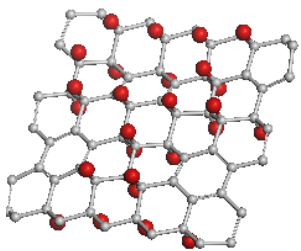 H26 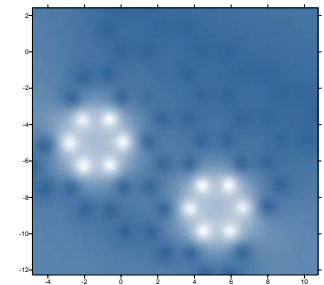

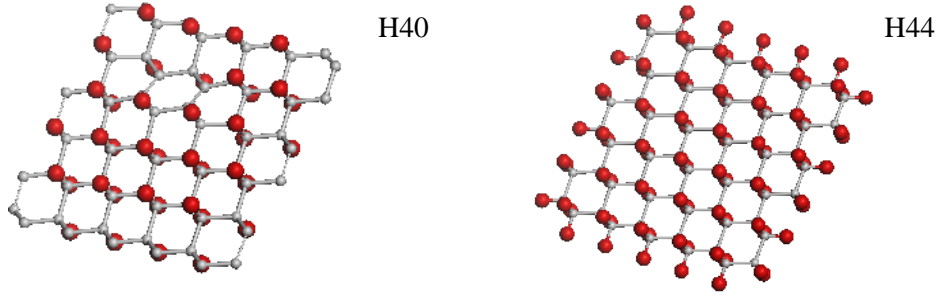

**Figure 4.** Equilibrium structure (left) and real-space ACS maps (right) of hydrides from the conclusive stage of the basal-plane hydrogenation.

$$E_{cpl}^{step}(n) = \Delta H_{nHgr} - \Delta H_{(n-1)Hgr}. \tag{2}$$

Evidently, both total and perstep coupling energies as well as the total energy of hydrides $\Delta H_{nHgr}$ are due to both the deformation of the nanographene carbon skeleton and the covalent coupling of hydrogen atoms with the substrate resulted in the formation of C-H bonds. Supposing that the relevant contributions can be summed up, one may try to evaluate them separately. Thus, the total deformation energy can be determined as the difference

$$E_{def}^{tot}(n) = \Delta H_{nHgr}^{sk} - \Delta H_{gr}. \tag{3}$$

Here $\Delta H_{nHgr}^{sk}$ determines the heat of formation of the carbon skeleton of the studied hydride at each stage of the hydrogenation. The value can be obtained as a result of one-point-structure determination of the equilibrium hydride after removing all hydrogen atoms. The deformation energy that accompanies each step of the hydrogenation can be determined as

$$E_{def}^{step}(n) = \Delta H_{nHgr}^{sk} - \Delta H_{(n-1)Hgr}^{sk}, \tag{4}$$

where $\Delta H_{nHgr}^{sk}$ and $\Delta H_{(n-1)Hgr}^{sk}$ match heats of formation of the carbon skeletons of the relevant hydrides at two subsequent steps of hydrogenation. Similarly, the total and partial chemical contributions caused by the formation of C-H bonds can be determined as

$$E_{cov}^{tot}(n) = E_{cpl}^{tot}(n) - E_{def}^{tot}(n) \tag{5}$$

and

$$E_{cov}^{step}(n) = E_{cpl}^{step}(n) - E_{def}^{step}(n-1). \tag{6}$$

Figure 6 displays the discussed calculated energies for a complete set of hydrides formed on the basis of (5, 5) nanographene fixed membrane. Zero points on horizontal axes correspond to the 44 atoms-framed membrane with empty basal plane. As seen in figure 6a, the total hydride energy $\Delta H_{nHgr}$ actually presents the difference of two big numbers whose absolute values increase when the number of hydrogen atoms grows. The difference is positive, while gradually decreasing by value, up

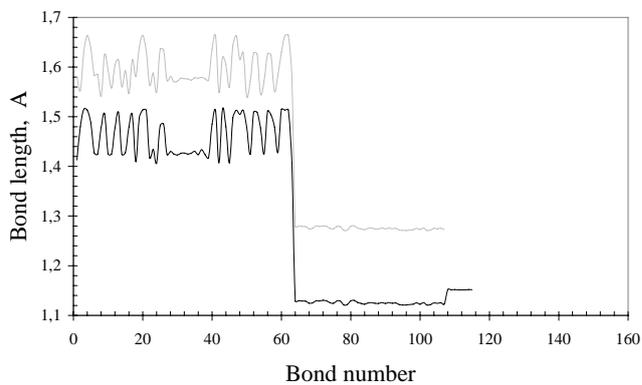

H8

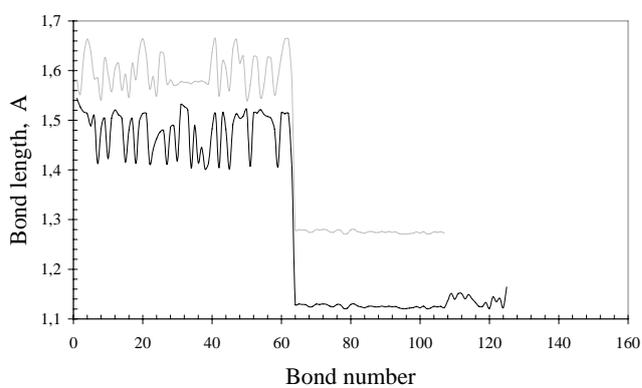

H18

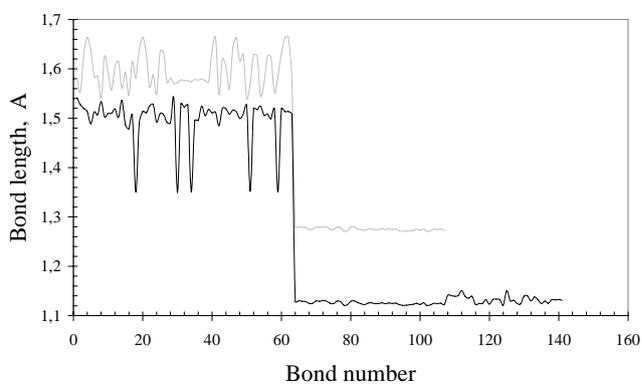

H34

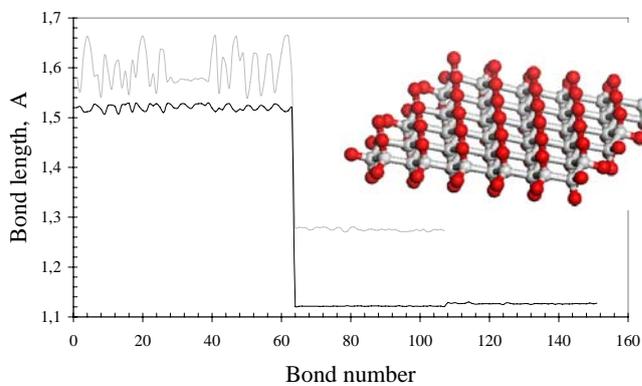

H44

**Figure 5**. $sp^2$-$sp^3$ Transformation of the graphene structure in due course of successive hydrogenation. Gray and black curves correspond to the bond length distribution related to the pristine graphene and its hydrides, respectively.

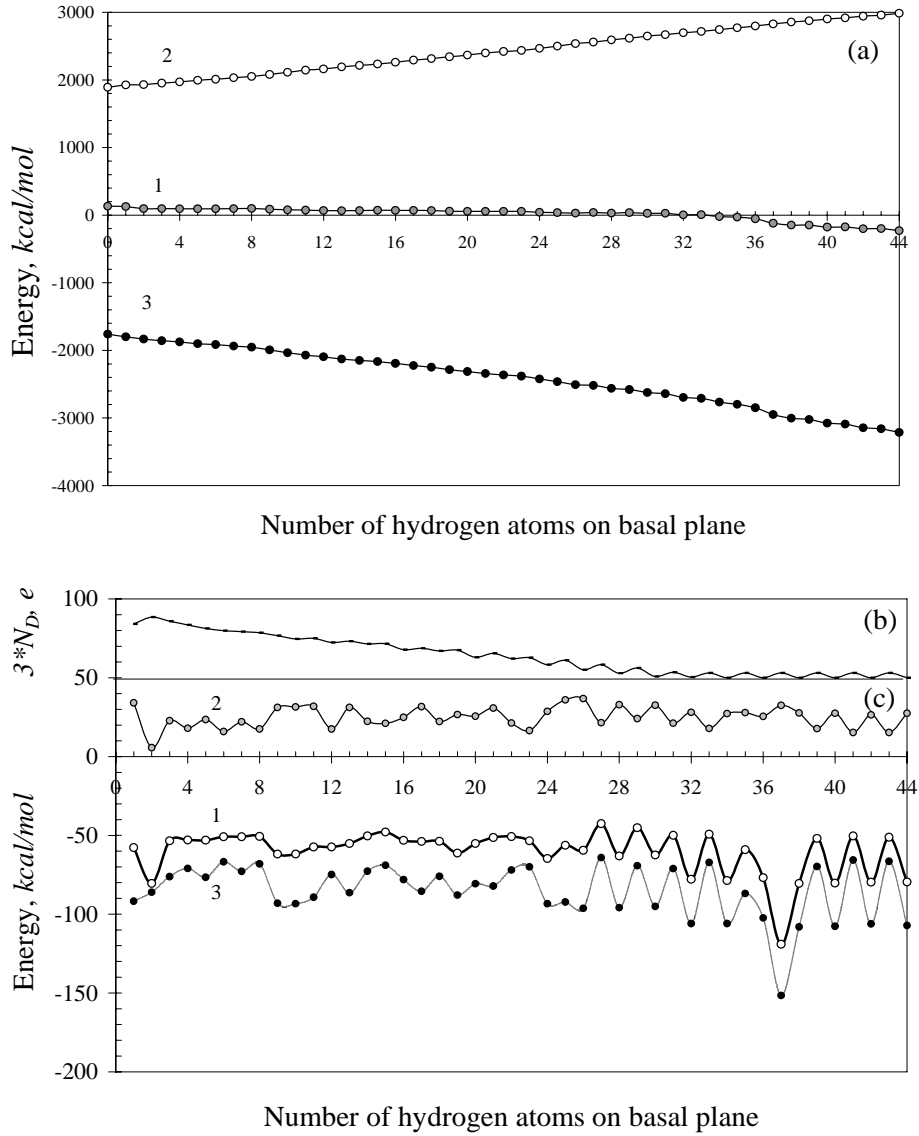

**Figure 6.** Energetic and electron characteristics of the graphene hydrides. a. Total energy $\Delta H_{(n-44)Hgr}$ (1); $E_{def}^{tot}(n-44)$ (2) and $E_{cov}^{tot}(n-44)$ (3) (see Eqs. (3) and (5)). b. The total number of effectively unpaired electrons. c. Per-step energies $E_{cpl}^{step}(n-44)$ (1); $E_{def}^{step}(n-44)$ (2); $E_{cov}^{step}(n-44)$ (3).

to the 32$^{th}$ step exhibiting a dominating contribution of deformational energy into the total one. At the 32$^{th}$ step $\Delta H_{nHgr}$ changes sign thus showing that the chemical contribution becomes dominating. Per-step energies in figure 6c clearly demonstrate that the hydrogenation is energetically profitable reaction until the end when all basal carbon atoms are involved in the process. A detail behavior of the deformation energy and the chemical contribution may be used for a scrupulous consideration of each hydrogenation step.

The efficacy of chemical reactions aimed at the formation of polyderivatives of $sp^2$ nanocarbons can be well characterized by the behavior of the total number of their effectively unpaired electrons, $N_D$, in due course of the reaction [53]. Initial value $N_D$ of the studied (5,5) nanographene molecule constitutes 31.04 *e*. After framing edge atoms by 44 hydrogens the value decreases up to 13.76 *e* and a gradual working out of this pool of the molecule chemical susceptibility in due course of the hydrogenation of the molecule basal plane is presented in figure 6b. To suit the scale of figure 6c, the $N_D$ data are multiplied by 3 and shifted up by 50 units. As seen in figure 6b, the hydrogenation is accompanied by gradual decreasing of the value until the latter reaches zero at the $32^{th}$ step. The further hydrogenation process was discussed in Section 5. Generally, the disclosed reaction is fully similar to that occurring when fullerene $C_{60}$ is hydrogenated [51, 52].

## 7. Discussion and conclusive remarks

Graphene hydrogenation is a complex process, the configuration of whose final products depend on many factors. Applying a molecular approach to the problem, it becomes possible to elucidate some of them, if not all. The undertaken investigation, part of whose results is presented in the current paper, seems to be a quite convincing demonstration of the approach ability. The obtained results have made possible to answer questions put earlier. However, before answering the questions we should concern the reason of chemical activity of graphene molecules as well as the driving force of their hydrogenation.

The partial radicalization of the graphene molecule is caused by its effectively unpaired odd electrons and is responsible for a quite efficient chemical activity of the body. The radicalization is a common feature for all benzenoid-patterned *sp²* nanocarbons, such as fullerenes, nanotubes, and graphenes, and is connected with exceeding C-C bond length in the bodies a critical value under which only a complete covalent binding of odd electrons occurs transforming them into classical π electrons [53]. The total number of effectively unpaired electrons $N_D$ determines the molecule chemical susceptibility and its gradual working out in due course of subsequent reaction steps governs the reaction pathway and provides its termination. In view of this general statement, it is possible to answer questions put in section 2.

1. Which kind of the hydrogen adsorption, namely, molecular or atomic, is the most probable?

Our study has convincingly shown that only atomic adsorption is effective and energetically favorable. In contrast with about 100% saturation of graphene atoms with individual hydrogen atoms, the deposition of the first hydrogen molecule on the basal plane of fixed membrane is not possible at all. All approaching hydrogens are not allowed to be accommodated on the plane and are repelled from it. In the case of free standing (5,5) nanographene membrane, the first hydrogen molecule meets difficulty and its adsorption is accompanied with positive coupling energy of 9 kcal/mol thus disclosing the energetic nonprofitability of the deposition. The second molecule is repelled from the plane at any way of approaching. Addition of the third molecule provides the deposition of the preceding one together with a new coming one. The total coupling energy is of 4 kcal/mol, which indicates to very weak and energetically non-profitable coupling with substrate. These and other difficulties accompany the deposition of next molecules. It seems that the reason in so dramatic difference between atomic and molecular adsorption is in the tendency of the graphene substrate to conserve the hexagon pattern. But obviously, the pattern conservation can be achieved if only the substrate hydrogenation provides the creation of cyclohexane-patterned structure in view of one of the conformers of the latter, or of their mixture. If non-coordinated deposition of individual atoms, as we saw in figure3 and figure 4, can meet the requirement in the case of individual atom deposition, a coordinated deposition of two atoms on neighboring carbons of substrate evidently make the formation of a cyclohexane-conformer pattern much less probable.

2. What is a characteristic image of the hydrogen atom attachment to the substrate?

The hydrogen atom is deposited on-top of the carbon ones in both up and down configurations. In contrast to a vast number of organic molecules, the length of C-H bonds formed under adsorption exceed 1.1Å, therewith differently for differently deposited atoms. Thus, C-H bonds are quite constant by value of 1.125Å in average for framing hydrogens that saturate edge carbon atoms of the substrate. Deposition on the basal plane causes enlarging the value up to maximum 1.152Å for H8. However, the formation of a regular chairlike cyclohexane structure like graphane leads to equalizing and shortening the bonds to 1.126Å. The above picture, which is characteristic for the fixed membrane, is significantly violated when going to one-side deposited fixed membrane or two-side deposited free standing membrane which exhibits the difference in the strength of the hydrogen atoms coupling with the related substrates.

3. Which carbon atom (or atoms) is the first target subjected to the hydrogen attachment? And
4. How carbon atoms are selected for the next steps of the adsorption?

Similarly to fullerenes and carbon nanotubes [53], the formation of graphene polyhydride $(CH)_n$ can be considered in the framework of stepwise computational synthesis, each subsequent step of which is controlled by the distribution of atomic chemical susceptibility in terms of partial numbers of effectively unpaired electrons on atom, $N_{DA}$, of preceding derivative over the substrate atoms. The high-rank $N_{DA}$ values definitely distinguish the atoms that should serve as targets for the next chemical attack. The successful generation of the polyderivatives families of fluorides and hydrides as well as other polyderivatives of fullerene $C_{60}$ [53], of polyhydride $(CH)_n$ related to chairlike graphane described in the current paper, of tablelike-cyclohexane graphane $(CH)_n$ [62] and chairlike graphene polyfluoride $(CF)_n$ [63] has shown a high efficacy of the approach in viewing the process of the polyderivatives formation which makes it possible to proceed with a deep insight into the mechanism of the chemical transformations.

5. Is there any connection between the sequential adsorption pattern and cyclohexane conformers?

The performed investigations have shown that there is a direct connection between the state of graphene substrate and the conformer pattern of the polyhydride formed. The pattern is governed by the cyclohexane conformer which formation under ambient conditions is the most profitable. Thus, a regular chairlike-cyclohexane conformed graphene with 100% hydrogen covering, known as graphane, is formed in the case when the graphene substrate is a perimeter-fixed membrane, both sides of which are accessible for hydrogen atoms.

When the membrane is two-side accessible but its edges are not fixed, the formation of a mixture of chairlike and boatlike cyclohexane patterns has turned out more profitable. As shown in the current paper, the polyhydride total energy involves deformational and chemical components. That is why the difference in the conformer energy in favor of chairlike conformer formed in free standing membrane is compensated by the gain in the deformation energy of the carbon carcass caused by the formation of boatlike conformer, which simulates a significant corrugation of the initial graphene plane [64]. The mixture of the two conformers transforms therewith a regular crystalline behavior of graphane into an amorphous-like behavior in the latter case.

When the membrane is one-side accessible but periphery-fixed, each hydrogenation step is accompanied by the energy growth and is energetically non-favorable [62]. The configuration produced is rather regular and looks like an infinite array of *trans*-linked tablelike cyclohexane conformers. The very fact that the tablelike conformer energy significantly exceeds that one for chairlike and boatlike conformers lays the foundation of the graphene hydride energy growth when the hydrogenation proceeds. The coupling of hydrogen atoms with the carbon skeleton is the weakest among all the considered configurations, which is particularly characterized by the longest C-H bonds of 1.18-1.21Å in length. The carbon skeleton takes a shape of a delta-plan canopy exterior therewith. However, under particular conditions that accompany, for example, the mineral shungit formation under specific hydrothermal conditions characterized by high temperature and pressure as well as high concentration of hydrogen atoms, this one-side hydrogenation may occur followed by spontaneous removing of weakly coupled hydrogen adsorbate from the carbon skeleton. The latter, looking like a concave saucer, forms particular carbon shells, whose compressing under high pressure may lead to the shungit formation [65].

**Acknowledgement**s
 The author (E.Sh) greatly appreciates a fruitful discussion with Prof. K.Novoselov.